\documentclass[prb,preprintnumbers,amsmath,amssymb,floatfix]{revtex4}

\usepackage{graphicx}
\usepackage{subfigure}
\usepackage{dcolumn}

\begin{document}

\title{Quantum theory of photonic crystals}
\author{Xiang-Yao Wu$^{a}$ \footnote{E-mail: wuxy2066@163.com},
 Ji Ma$^{a}$, Xiao-Jing Liu$^{a}$, Jing-Hai Yang$^{a}$\\ Hong Li$^{a}$,
  Si-Qi Zhang$^{a}$, Hai-Xin Gao$^{b}$, Xin-Guo Yin$^{c}$ and San Chen$^{c}$}
 \affiliation{a. Institute of Physics, Jilin Normal
University,
Siping 136000 China\\
b. Institute of Physics, Northeast Normal University, Changchun
130024 China\\
c. Institute of Physics, Huaibei Normal University, Huaibei 235000
China }
%%%%%%%
\begin{abstract}
In this paper, we have firstly presented a new quantum theory to
study one-dimensional photonic crystals. We give the quantum
transform matrix, quantum dispersion relation and quantum
transmissivity, and compare them with the classical dispersion
relation and classical transmissivity. By the calculation, we find
the classical and quantum dispersion relation and transmissivity
are identical. The new approach can be studied two-dimensional and
three-dimensional photonic crystals.\\
\\
\vskip 5pt

PACS: 42.70.Qs, 78.20.Ci, 41.20.Jb\\
Keywords:  Photonic crystals; Quantum transmissivity; Quantum
dispersion relation
\end{abstract}
\maketitle
{\bf 1. Introduction} \vskip 8pt

Photonic crystals (PCs) are artificial materials with periodic
variations in refractive index that are designed to affect the
propagation of light [1-4]. An important feature of the PCs is
that there are allowed and forbidden ranges of frequencies at
which light propagates in the direction of index periodicity. Due
to the forbidden frequency range, known as photonic band gap (PBG)
[5-6], which forbids the radiation propagation in a specific range
of frequencies. The existence of PBGs will lead to many
interesting phenomena. In the past ten years has been developed an
intensive effort to study and micro-fabricate PBG materials in
one, two or three dimensions, e.g., modification of spontaneous
emission [7-9] and photon localization [10-14].

Thus numerous applications of PCs have been proposed in improving
the performance of optoelectronic and microwave devices such as
high-efficiency semiconductor lasers, right emitting diodes, wave
guides, optical filters, high-Q resonators, antennas,
frequency-selective surface, optical wave guides and sharp bends
[15], WDM-devices [16-17], splitters and combiners [18]. optical
limiters and amplifiers [19-20].

At present, the theory calculations of PCs have many numerical
methods, such as: the plane-wave expansion method (PWE) [21-23],
the finite-difference time-domain method (FDTD) [24-27], the
transfer matrix method (TMM) [28-29], the finite element method
(FE) [30-33], the scattering matrix method [34], the Green's
function method [35] and so on. These methods are classical
electromagnetism theory. Obviously, the full quantum theory of PCs
is necessary. In Refs. [36-37], the authors give the quantum wave
equation of single photon. In Ref. [38], we give the quantum wave
equations of free and non-free photon. In this paper, We have
studied the 1D PCs by the quantum wave equations of photon [38],
and give quantum dispersion relation, quantum transmissivity and
reflectivity, and obtain some new results, which can be tested by
experiments. Obviously, the new method of quantum theory can be
studied the 2D and 3D PCs.

\vskip 8pt

{\bf 2. The quantum wave equation and probability current density
of photon} \vskip 8pt

The quantum wave equations of free and non-free photon have been
obtained in Ref. [39], they are
\begin{equation}
i\hbar\frac{\partial}{\partial
t}\vec{\psi}(\vec{r},t)=c\hbar\nabla\times\vec{\psi}(\vec{r},t),
\end{equation}
and
\begin{equation}
i\hbar\frac{\partial}{\partial
t}\vec{\psi}(\vec{r},t)=c\hbar\nabla\times\vec{\psi}(\vec{r},t)+V\vec{\psi}(\vec{r},t),
\end{equation}
where $\vec{\psi}(\vec{r},t)$ is the vector wave function of
photon, and $V$ is the potential energy of photon in medium. In
the medium of refractive index $n$, the photon's potential energy
$V$ is [39]
\begin{equation}
V=\hbar\omega(1-n).
\end{equation}
The conjugate of Eq. (2) is
\begin{equation}
-i\hbar\frac{\partial}{\partial
t}\vec{\psi}^{\ast}(\vec{r},t)=c\hbar\nabla\times\vec{\psi}^{\ast}(\vec{r},t)+V\vec{\psi}^{\ast}(\vec{r},t).
\end{equation}
Multiplying the Eq. (2) by $\vec{\psi}^{\ast}$, the Eq. (4) by
$\vec{\psi}$, and taking the difference, we get
\begin{equation}
i\hbar\frac{\partial}{\partial
t}(\vec{\psi}^{\ast}\cdot\vec{\psi})=c\hbar(\vec{\psi}^{\ast}\cdot\nabla\times\vec{\psi}-
\vec{\psi}\cdot\nabla\times\vec{\psi}^{\ast})=c\hbar\nabla\cdot(\vec{\psi}\times\vec{\psi}^{\ast}),
\end{equation}
i.e.
\begin{equation}
\frac{\partial\rho}{\partial t}+\nabla\cdot J=0,
\end{equation}
where
\begin{equation}
\rho=\vec{\psi}^{\ast}\cdot\vec{\psi},
\end{equation}
and
\begin{equation}
J=ic\vec{\psi}\times\vec{\psi}^{\ast},
\end{equation}
are the probability density and probability current density,
respectively.

By the method of separation variable
\begin{equation}
\vec{\psi}(\vec{r},t)=\vec{\psi}(\vec{r})f(t),
\end{equation}
the time-dependent Eq. (2) becomes the time-independent equation
\begin{equation}
c\hbar\nabla\times\vec{\psi}(\vec{r})+V\vec{\psi}(\vec{r})=E\vec{\psi}(\vec{r}),
\end{equation}
where $E$ is the energy of photon in medium.

By taking curl in (10), when $\frac{\partial V}{\partial x_{i}}=0,
 (i=1, 2, 3)$, the Eq. (10) becomes
\begin{equation}
(\hbar
c)^{2}(\nabla(\nabla\cdot\vec{\psi}(\vec{r}))-\nabla^2\vec{\psi}(\vec{r}))=(E-V)^2\vec{\psi}(\vec{r}).
\end{equation}
Choosing transverse gange $\nabla\cdot\vec{\psi}(\vec{r})=0$, Eq.
(11) becomes
\begin{equation}
\nabla^2\vec{\psi}(\vec{r})+(\frac{E-V}{\hbar
c})^2\vec{\psi}(\vec{r})=0.
\end{equation}
In vacuum, potential energy $V=0$, Eq. (12) becomes
\begin{equation}
\nabla^2\vec{\psi}(\vec{r})+K^2\vec{\psi}(\vec{r})=0.
\end{equation}
Where $K=\frac{\omega}{c}$. Eqs. (12) and (13) are the quantum
wave equation of photon in medium and vacuum, and we can study
one-dimensional PCs by them.

\vskip 8pt

{\bf 3. The quantum theory of one-dimensional Photonic crystals}
\vskip 8pt

For one-dimensional Photonic crystals, we should define and
calculate its quantum dispersion relation and quantum
transmissivity. The one-dimensional PCs structure is shown in FIG.
1.

In FIG. 1, $\vec{\psi}_{I}$, $\vec{\psi}_{R}$, $\vec{\psi}_{T}$
are the wave functions of incident, reflection and transmission
photon, respectively. By Eq. (13), they can be written as
\begin{figure}[tbp]
\includegraphics[width=9 cm]{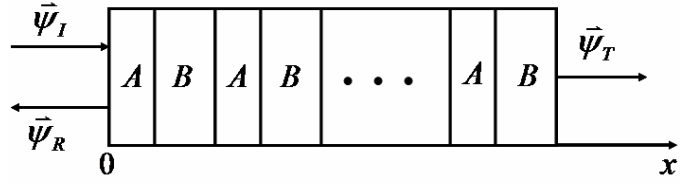}
\caption{the structure of one-dimensional photonic crystals}
\end{figure}

\begin{eqnarray}
\vec{\psi}(\vec{r},t)&=&\vec{\psi}_{0}e^{i(\vec{K}\cdot\vec{r}-\omega
t)}=\psi_{x}\vec{i}+\psi_{y}\vec{j}+\psi_{z}\vec{k}\nonumber\\
&=&\psi_{x0}e^{i(\vec{K}\cdot\vec{r}-\omega
t)}\vec{i}+\psi_{y0}e^{i(\vec{K}\cdot\vec{r}-\omega
t)}\vec{j}+\psi_{z0}e^{i(\vec{K}\cdot\vec{r}-\omega t)}\vec{k},
\end{eqnarray}
By transverse gange $\nabla\cdot\vec{\psi}(\vec{r})=0$, we get
\begin{equation}
K_{x}\psi_{x}+K_{y}\psi_{y}+K_{z}\psi_{z}=0.
\end{equation}
In FIG. 1, the photon travels along with the $x$ axis, the wave
vector $K_{y}=K_{z}=0$ and $K_{x}\neq 0$. By Eq. (15), we have
\begin{equation}
\psi_{x}=0,
\end{equation}
so the total wave function of photon is
\begin{equation}
\vec{\psi}=\vec{\psi}_y\vec{j}+\vec{\psi}_z\vec{k}.
\end{equation}
By the wave function continuum, the $\psi_{x}=0$ in medium. So,
the Eq. (13) becomes two component equations
\begin{equation}
\nabla^2\psi_y+(\frac{E-V}{\hbar c})^2\psi_y=0,
\end{equation}
and
\begin{equation}
\nabla^2\psi_z+(\frac{E-V}{\hbar c})^2\psi_z=0.
\end{equation}
In FIG. 1, the wave functions of incident, reflection and
transmission photon can be written as
\begin{equation}
\vec{\psi_{I}}=F_{y}e^{i(\vec{K}\cdot\vec{r}-\omega
t)}\vec{j}+F_{z}e^{i(\vec{K}\cdot\vec{r}-\omega t)}\vec{k},
\end{equation}
\begin{equation}
\vec{\psi_{R}}=F_{y}^{'}e^{-i(\vec{K}\cdot\vec{r}+\omega
t)}\vec{j}+F_{z}^{'}e^{-i(\vec{K}\cdot\vec{r}+\omega t)}\vec{k},
\end{equation}
\begin{equation}
\vec{\psi_{T}}=D_{y}e^{i(\vec{K}\cdot\vec{r}-\omega
t)}\vec{j}+D_{z}e^{i(\vec{K}\cdot\vec{r}-\omega t)}\vec{k},
\end{equation}
where $F_{y}$, $F_{z}$, $F_{y}^{'}$, $F_{z}^{'}$, $D_{y}$, and
$D_{z}$ are their amplitudes.\\

The component form of Eq. (1) is
\begin{eqnarray}
\left \{ \begin{array}{lll} && i\hbar\frac{\partial}{\partial
t}\psi_x=\hbar c(\frac{\partial \psi_z}{\partial y}-\frac{\partial
\psi_y}{\partial z})\\&& i\hbar\frac{\partial}{\partial
t}\psi_y=\hbar c(\frac{\partial \psi_x}{\partial z}-\frac{\partial
\psi_z}{\partial x})
\\&& i\hbar\frac{\partial}{\partial
t}\psi_z=\hbar c(\frac{\partial \psi_y}{\partial x}-\frac{\partial
\psi_x}{\partial y})
    \end{array}
   \right.,
\end{eqnarray}
substituting Eqs. (14) and (16) into (23), we have
\begin{equation}
\psi_z=i\psi_y,
\end{equation}
the probability current density becomes
\begin{equation}
J=ic\vec{\psi}\times\vec{\psi}^{\ast}=2c|\psi_z|^{2}\vec{i}=2c|\psi_{0z}|^{2}\vec{i},
\end{equation}
where
\begin{equation}
\psi_z=\psi_{0z}e^{i(\vec{k}\cdot\vec{r}-\omega t)},
\end{equation}
the $\psi_{0z}$ is $\psi_z$ amplitude.\\

For the incident,reflection and transmission photon, their
probability current density $J_I$, $J_R$, $J_T$ are
\begin{equation}
J_I=2c|F_z|^{2},
\end{equation}
\begin{equation}
J_R=2c|F_z^{'}|^{2},
\end{equation}
\begin{equation}
J_T=2c|D_z|^{2},
\end{equation}
We can define quantum transmissivity $T$ and quantum reflectivity
$R$ as
\begin{equation}
T=\frac{J_{T}}{J_{I}}=|\frac{D_z}{F_z}|^2,
\end{equation}
\begin{equation}
R=\frac{J_{R}}{J_{I}}=|\frac{F'_z}{F_z}|^2.
\end{equation}
With Eqs. (30) and (31), we find quantum transmissivity and
reflectivity are relevant to the $z$ component amplitudes of wave
function of the incident, reflection and transmission photon.

 \vskip 8pt

{\bf 4. The quantum transmissivity and quantum dispersion
relation} \vskip 8pt

Since the quantum transmissivity is relevant to the $z$ component
amplitude of transmission wave function, we should only solve the
$z$ component equation (19) for the one-dimensional PCs, which is
shown in FIG. 2

\begin{figure}[tbp]
\includegraphics[width=9 cm]{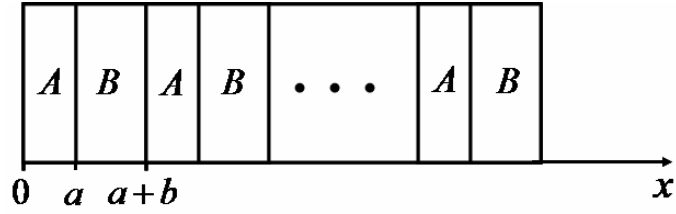}
\caption{the structure of one-dimensional photonic crystals}
\end{figure}

With Eq. (19), the photon's quantum wave equation in mediums $A$
and $B$ are

\begin{equation}
\frac{\partial^{2}\psi_A}{\partial
x^{2}}+k_{A}^2\psi_A=0\hspace{0.6in}(0<x<a),
\end{equation}
\begin{equation}
\frac{\partial^{2}\psi_B}{\partial
x^{2}}+k_{B}^2\psi_B=0\hspace{0.6in}(a<x<a+b),
\end{equation}
where
\begin{equation}
k_{A}=\frac{E-V_a}{\hbar c}=\frac{E-\hbar\omega(1-n_{a})}{\hbar
c}=\frac{\omega}{c}n_{a}=\frac{2\pi}{\lambda}n_{a},
\end{equation}
\begin{equation}
k_{B}=\frac{E-V_b}{\hbar c}=\frac{E-\hbar\omega(1-n_{b})}{\hbar
c}=\frac{\omega}{c}n_{b}=\frac{2\pi}{\lambda}n_{b},
\end{equation}
where $\lambda={2\pi c}/{\omega}$ is the photon wave length in
vacuum, $V_a=\hbar\omega(1-n_{a})
\hspace{0.05in}(V_b=\hbar\omega(1-n_{b}))$ is the potential energy
of photon in medium $A (B)$, and $n_{a} (n_{b})$ is refractive
index of medium $A (B)$. In order to simplify, the index $z$ is
omitted, i.e., $\psi_{zA} (\psi_{zB})$ is written as $\psi_{A}
(\psi_{B})$.
 \\
The solutions of Eqs. (32) and (33) are
\begin{equation}
{\psi_A}=A_{1}e^{ik_Ax}+A_{2}e^{-ik_Ax}\hspace{0.6in}(0<x<a),
\end{equation}
\begin{equation}
{\psi_B}=B_{1}e^{ik_Bx}+B_{2}e^{-ik_Bx}\hspace{0.6in}(a<x<a+b).
\end{equation}
By Bloch law, there is
\begin{eqnarray}
{\psi(a+b<x<2a+b)}&=&\psi(0<x<a)e^{ik(a+b)} \nonumber\\
&=&(A_{1}e^{ik_A(x-(a+b))}\nonumber\\&&+A_{2}e^{-ik_A(x-(a+b))})e^{ik(a+b)},
\end{eqnarray}
where $k$ is Bloch wave vector. \\
At $x=a$, by the continuation of wave function and its derivative,
we have
\begin{eqnarray}
 A_{1}e^{ik_Aa}+A_{2}e^{-ik_Aa}=B_{1}e^{ik_Ba}+B_{2}e^{-ik_Ba},
\end{eqnarray}
\begin{eqnarray}
 ik_AA_{1}e^{ik_Aa}-ik_AA_{2}e^{-ik_Aa}=ik_BB_{1}e^{ik_Ba}-ik_BB_{2}e^{-ik_Ba},
\end{eqnarray}
At $x=a+b$, by the continuation of wave function and its
derivative, we have
\begin{eqnarray}
 A_{1}e^{i k(a+b)}+A_{2}e^{i
 k(a+b)}=B_{1}e^{ik_B(a+b)}+B_{2}e^{-ik_B(a+b)},
\end{eqnarray}
\begin{eqnarray}
 ik_AA_{1}e^{i k(a+b)}-ik_AA_{2}e^{i
 k(a+b)}=ik_BB_{1}e^{ik_B(a+b)}-ik_BB_{2}e^{-ik_B(a+b)},
\end{eqnarray}
and we obtain the follows equations set
\begin{eqnarray}
\left \{ \begin{array}{ll}
    A_{1}e^{ik_Aa}+A_{2}e^{-ik_Aa}=B_{1}e^{ik_Ba}+B_{2}e^{-ik_Ba} \\
  ik_AA_{1}e^{ik_Aa}-ik_AA_{2}e^{-ik_Aa}=ik_BB_{1}e^{ik_Ba}-ik_BB_{2}e^{-ik_Ba} \\
  A_{1}e^{i k(a+b)}+A_{2}e^{i k(a+b)}=B_{1}e^{ik_B(a+b)}+B_{2}e^{-ik_B(a+b)} \\
   ik_AA_{1}e^{i k(a+b)}-ik_AA_{2}e^{i k(a+b)}=ik_BB_{1}e^{ik_B(a+b)}-ik_BB_{2}e^{-ik_B(a+b)},\\
   \end{array}
   \right.
\end{eqnarray}
the necessary and sufficient condition of Eq. (43) nonzero
solution is its coefficient determinant equal to zero
\begin{eqnarray}
&&\left | \begin{array}{llll}
 \hspace{0.2in}e^{ik_Aa}   \hspace{0.6in}e^{-ik_Aa}  \hspace{0.6in}-e^{ik_Ba}   \hspace{0.5in}-e^{-ik_Ba} \\
 \hspace{0.2in}k_Ae^{ik_Aa}   \hspace{0.4in}-k_Ae^{-ik_Aa}   \hspace{0.3in}-k_Be^{ik_Ba} \hspace{0.4in}k_Be^{-ik_Ba} \\
 \hspace{0.2in}e^{ik(a+b)}   \hspace{0.4in}e^{ik(a+b)}  \hspace{0.6in}-e^{ik_B(a+b)}   \hspace{0.3in}-e^{-ik_B(a+b)} \\
 \hspace{0.2in}k_Ae^{ik(a+b)}  \hspace{0.2in}-k_Ae^{ik(a+b)}   \hspace{0.3in}-k_Be^{ik_B(a+b)}  \hspace{0.2in}k_Be^{-ik_B(a+b)} \\

   \end{array}
   \right |=0,
   \end{eqnarray}
simplifying Eq. (44), we obtain the quantum dispersion relation
\begin{eqnarray}
\cos(k(a+b))=\cos(k_Aa)\cos(k_Bb)-\frac{1}{2}(\frac{1}{k_A}+\frac{1}{k_B})\sin(k_Aa)\sin(k_Bb).
\end{eqnarray}
In the following, we should give the wave function of photon in
every medium, and the transmission wave function. In FIG. 3, we
give the simplification form of wave function in every medium,
such as symbols $A_{k_A}^{1}$ and $A_{-k_A}^{1}$ express
simplifying wave function of medium $A$ in the first period, it
express wave function
\begin{equation}
{\psi_{A^{1}}(x)}=A_{k_A}^{1}e^{ik_Ax}+A_{-k_A}^{1}e^{-ik_Ax},
\end{equation}
in medium $B$ of first period, the symbols $B_{k_A}^{1}$ and
$B_{-k_A}^{1}$ express wave function
\begin{equation}
{\psi_{B^{1}}(x)}=B_{k_B}^{1}e^{ik_Bx}+B_{-k_B}^{1}e^{-ik_Bx},
\end{equation}
in medium $A$ of second period, the symbols $A_{k_A}^{2}$ and
$A_{-k_A}^{2}$ express wave function
\begin{equation}
{\psi_{A^{2}}(x)}=A_{k_A}^{2}e^{ik_Ax}+A_{-k_A}^{2}e^{-ik_Ax},
\end{equation}
similarly, in medium $B$ of second period, the symbols
$B_{k_A}^{2}$ and $B_{-k_A}^{2}$ express wave function
\begin{equation}
{\psi_{B^{2}}(x)}=B_{k_B}^{2}e^{ik_Bx}+B_{-k_B}^{2}e^{-ik_Bx},
\end{equation}
and so on.\\
In the incident area, the total wave function $\psi_{tot}(x)$ is
the superposition of incident and reflection wave function, it is
\begin{figure}[tbp]
\includegraphics[width=12 cm]{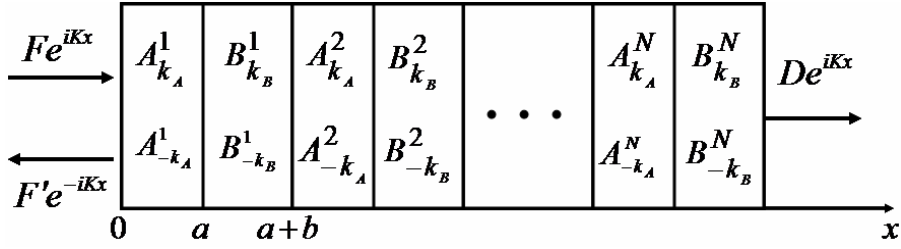}
\caption{the quantum structure of one-dimensional photonic
crystals}
\end{figure}
\begin{equation}
{\psi_{tot}(x)}={\psi_{I}(x)}+{\psi_{R}(x)}=Fe^{iKx}+F'e^{-iKx},
\end{equation}

where $K$ is the wave vector of incident, reflection, and
transmission photon. In the following, we should use the condition
of wave function and its
derivative continuation at interface of two mediums.\\
(1) At $x=0$, by the continuation of wave function and its
derivative, we have
\begin{equation}
F+F'=A_{k_A}^{1}+A_{-k_A}^{1},
\end{equation}
\begin{equation}
iKF-iKF'=ik_AA_{k_A}^{1}-ik_AA_{-k_A}^{1},
\end{equation}
we obtain
\begin{equation}
A_{k_A}^{1}=\frac{1}{2}[(1+\frac{K}{k_A})F+(1-\frac{K}{k_A})F'],
\end{equation}
\begin{equation}
A_{-k_A}^{1}=\frac{1}{2}[(1-\frac{K}{k_A})F+(1+\frac{K}{k_A})F'],
\end{equation}
the Eqs. (53) and (54) can be written as matrix form
\begin{eqnarray}
\left ( \begin{array}{ll}
 A_{k_A}^{1}\\
 A_{-k_A}^{1}\\
   \end{array}
   \right )=\frac{1}{2} \left ( \begin{array}{ll}
  1+{K}/{k_A} \hspace{0.2in} 1-{K}/{k_A}\\
 1-{K}/{k_A} \hspace{0.2in} 1+{K}/{k_A}\\

   \end{array}
   \right )\left ( \begin{array}{ll}
F\\
F'\\
   \end{array}
   \right )=M_A^1\left ( \begin{array}{ll}
F\\
F'\\
   \end{array}
   \right ),
\end{eqnarray}
where $M_A^1$ is the quantum transform matrix of the first period
medium $A$, it is
\begin{eqnarray}
M_A^1=\frac{1}{2} \left ( \begin{array}{ll}
  1+{K}/{k_A} \hspace{0.2in} 1-{K}/{k_A}\\
 1-{K}/{k_A} \hspace{0.2in} 1+{K}/{k_A}\\

   \end{array}
   \right ),
\end{eqnarray}

(2) At $x=a$, by the continuation of wave function and its
derivative, we have
\begin{eqnarray}
 A_{k_A}^{1}e^{ik_Aa}+A_{-k_A}^{1}e^{-ik_Aa}=B_{k_B}^{1}e^{ik_Ba}+B_{-k_B}^{1}e^{-ik_Ba},
\end{eqnarray}
\begin{eqnarray}
 \frac{k_A}{k_B}(A_{k_A}^{1}e^{ik_Aa}-A_{-k_A}^{1}e^{-ik_Aa})=B_{k_B}^{1}e^{ik_Ba}-B_{-k_B}^{1}e^{-ik_Ba},
\end{eqnarray}
we get
\begin{eqnarray}
B_{k_B}^{1}=\frac{1}{2}e^{i(k_A-k_B)a}(1+\frac{k_A}{k_B})A_{k_A}^{1}+\frac{1}{2}e^{-i(k_A+k_B)a}(1-\frac{k_A}{k_B})A_{-k_A}^{1},
\end{eqnarray}
\begin{eqnarray}
B_{-k_B}^{1}=\frac{1}{2}e^{i(k_A+k_B)a}(1-\frac{k_A}{k_B})A_{k_A}^{1}+\frac{1}{2}e^{i(k_B-k_A)a}(1+\frac{k_A}{k_B})A_{-k_A}^{1},
\end{eqnarray}
the Eqs. (59) and (60) can be written as matrix form
\begin{eqnarray}
\left ( \begin{array}{ll}
B_{k_B}^{1}\\
 B_{-k_B}^{1}\\
   \end{array}
   \right )=\frac{1}{2} \left ( \begin{array}{ll}
  e^{i(k_A-k_B)a}(1+{k_A}/{k_B}) \hspace{0.2in} e^{-i(k_A+k_B)a}(1-{k_A}/{k_B})\\
e^{i(k_A+k_B)a}(1-{k_A}/{k_B}) \hspace{0.2in} e^{i(k_B-k_A)a}(1+{k_A}/{k_B})\\

   \end{array}
   \right )\left ( \begin{array}{ll}
A_{kA}^1\\
A_{-kA}^1\\
   \end{array}
   \right )=M_B^1\left ( \begin{array}{ll}
A_{kA}^1\\
A_{-kA}^1\\
   \end{array}
   \right ),
\end{eqnarray}
where $M_B^1$ is the quantum transform matrix of the first period
medium $B$, it is
\begin{eqnarray}
M_B^1=\frac{1}{2} \left ( \begin{array}{ll}
  e^{i(k_A-k_B)a}(1+{k_A}/{k_B}) \hspace{0.2in} e^{-i(k_A+k_B)a}(1-{k_A}/{k_B})\\
e^{i(k_A+k_B)a}(1-{k_A}/{k_B}) \hspace{0.2in} e^{i(k_B-k_A)a}(1+{k_A}/{k_B})\\

   \end{array}
   \right ),
\end{eqnarray}

(3) At $x=a+b$, by the continuation of wave function and its
derivative, we have
\begin{eqnarray}
B_{k_B}^{1}e^{ik_B(a+b)}+B_{-k_B}^{1}e^{-ik_B(a+b)}=A_{k_A}^{2}e^{ik_A(a+b)}+A_{-k_A}^{2}e^{-ik_A(a+b)},
\end{eqnarray}
\begin{eqnarray}
\frac{k_B}{k_A}(B_{k_B}^{1}e^{ik_B(a+b)}-B_{-k_B}^{1}e^{-ik_B(a+b)})=A_{k_A}^{2}e^{ik_A(a+b)}-A_{-k_A}^{2}e^{-ik_A(a+b)},
\end{eqnarray}
we get
\begin{eqnarray}
A_{k_A}^{2}=\frac{1}{2}e^{i(k_B-k_A)(a+b)}(1+\frac{k_B}{k_A})B_{k_B}^{1}+\frac{1}{2}e^{-i(k_A+k_B)(a+b)}(1-\frac{k_B}{k_A})B_{-k_B}^{1},
\end{eqnarray}
\begin{eqnarray}
A_{-k_A}^{2}=\frac{1}{2}e^{i(k_A+k_B)(a+b)}(1-\frac{k_B}{k_A})B_{k_B}^{1}+\frac{1}{2}e^{i(k_A-k_B)(a+b)}(1+\frac{k_B}{k_A})B_{-k_B}^{1},
\end{eqnarray}
the Eqs. (65) and (66) can be written as matrix form
\begin{eqnarray}
\left ( \begin{array}{ll}
A_{k_A}^{2}\\
 A_{-k_A}^{2}\\
   \end{array}
   \right )&=&\frac{1}{2} \left ( \begin{array}{ll}
  e^{i(k_B-k_A)(a+b)}(1+{k_B}/{k_A}) \hspace{0.1in} e^{-i(k_A+k_B)(a+b)}(1-{k_B}/{k_A})\\
e^{i(k_A+k_B)(a+b)}(1-{k_B}/{k_A}) \hspace{0.1in} e^{i(k_A-k_B)(a+b)}(1+{k_B}/{k_A})\\

   \end{array}
   \right )\left ( \begin{array}{ll}
B_{kB}^1\\
B_{-kB}^1\\
   \end{array}
   \right ) \nonumber\\&=&M_A^2\left ( \begin{array}{ll}
B_{kB}^1\\
B_{-kB}^1\\
   \end{array}
   \right ),
\end{eqnarray}
where $M_A^2$ is the quantum transform matrix of the second period
medium $A$, it is
\begin{eqnarray}
M_A^2=\frac{1}{2} \left ( \begin{array}{ll}
  e^{i(k_B-k_A)(a+b)}(1+{k_B}/{k_A}) \hspace{0.1in} e^{-i(k_A+k_B)(a+b)}(1-{k_B}/{k_A})\\
e^{i(k_A+k_B)(a+b)}(1-{k_B}/{k_A}) \hspace{0.1in} e^{i(k_A-k_B)(a+b)}(1+{k_B}/{k_A})\\

   \end{array}
   \right ),
\end{eqnarray}

(4) at $x=2a+b$, by the continuation of wave function and its
derivative, we get
\begin{eqnarray}
\left ( \begin{array}{ll}
B_{k_B}^{2}\\
 B_{-k_B}^{2}\\
   \end{array}
   \right )&=&\frac{1}{2} \left ( \begin{array}{ll}
  e^{i(k_A-k_B)(2a+b)}(1+{k_A}/{k_B}) \hspace{0.2in} e^{-i(k_A+k_B)(2a+b)}(1-{k_A}/{k_B})\\
e^{i(k_A+k_B)(2a+b)}(1-{k_A}/{k_B}) \hspace{0.2in} e^{i(k_B-k_A)(2a+b)}(1+{k_A}/{k_B})\\

   \end{array}
   \right )\left ( \begin{array}{ll}
A_{kA}^2\\
A_{-kA}^2\\
   \end{array}
   \right ) \nonumber\\&=&M_B^2\left ( \begin{array}{ll}
A_{kA}^2\\
A_{-kA}^2\\
   \end{array}
   \right ),
\end{eqnarray}
where $M_B^2$ is the quantum transform matrix of the second period
medium $B$, it is
\begin{eqnarray}
M_B^2=\frac{1}{2} \left ( \begin{array}{ll}
  e^{i(k_A-k_B)(2a+b)}(1+{k_A}/{k_B}) \hspace{0.2in} e^{-i(k_A+k_B)(2a+b)}(1-{k_A}/{k_B})\\
e^{i(k_A+k_B)(2a+b)}(1-{k_A}/{k_B}) \hspace{0.2in} e^{i(k_B-k_A)(2a+b)}(1+{k_A}/{k_B})\\

   \end{array}
   \right ),
\end{eqnarray}

(5) at $x=2(a+b)$, by the continuation of wave function and its
derivative, we get
\begin{eqnarray}
\left ( \begin{array}{ll}
A_{k_A}^{3}\\
 A_{-k_A}^{3}\\
   \end{array}
   \right )&=&\frac{1}{2} \left ( \begin{array}{ll}
  e^{i(k_B-k_A)2(a+b)}(1+{k_B}/{k_A}) \hspace{0.1in} e^{-i(k_A+k_B)2(a+b)}(1-{k_B}/{k_A})\\
e^{i(k_A+k_B)2(a+b)}(1-{k_B}/{k_A}) \hspace{0.1in} e^{i(k_A-k_B)2(a+b)}(1+{k_B}/{k_A})\\

   \end{array}
   \right )\left ( \begin{array}{ll}
B_{kB}^2\\
B_{-kB}^2\\
   \end{array}
   \right ) \nonumber\\&=&M_A^3\left ( \begin{array}{ll}
B_{kB}^2\\
B_{-kB}^2\\
   \end{array}
   \right ),
\end{eqnarray}
where $M_A^3$ is the quantum transform matrix of the third period
medium $A$, it is
\begin{eqnarray}
M_A^3=\frac{1}{2} \left ( \begin{array}{ll}
  e^{i(k_B-k_A)2(a+b)}(1+{k_B}/{k_A}) \hspace{0.1in} e^{-i(k_A+k_B)2(a+b)}(1-{k_B}/{k_A})\\
e^{i(k_A+k_B)2(a+b)}(1-{k_B}/{k_A}) \hspace{0.1in} e^{i(k_A-k_B)2(a+b)}(1+{k_B}/{k_A})\\

   \end{array}
   \right )
\end{eqnarray}
(6) similarly, at $x=3a+2b$, by the continuation of wave function
and its derivative, we get
\begin{eqnarray}
\left ( \begin{array}{ll}
B_{k_B}^{3}\\
 B_{-k_B}^{3}\\
   \end{array}
   \right )&=&\frac{1}{2} \left ( \begin{array}{ll}
  e^{i(k_A-k_B)(3a+2b)}(1+{k_A}/{k_B}) \hspace{0.2in} e^{-i(k_A+k_B)(3a+2b)}(1-{k_A}/{k_B})\\
e^{i(k_A+k_B)(3a+2b)}(1-{k_A}/{k_B}) \hspace{0.2in} e^{i(k_B-k_A)(3a+2b)}(1+{k_A}/{k_B})\\

   \end{array}
   \right )\left ( \begin{array}{ll}
A_{kA}^3\\
A_{-kA}^3\\
   \end{array}
   \right ) \nonumber\\&=&M_B^3\left ( \begin{array}{ll}
A_{kA}^3\\
A_{-kA}^3\\
   \end{array}
   \right ),
\end{eqnarray}
where $M_B^3$ is the quantum transform matrix of the third period
medium $B$, it is
\begin{eqnarray}
M_B^3=\frac{1}{2} \left ( \begin{array}{ll}
  e^{i(k_A-k_B)(3a+2b)}(1+{k_A}/{k_B}) \hspace{0.2in} e^{-i(k_A+k_B)(3a+2b)}(1-{k_A}/{k_B})\\
e^{i(k_A+k_B)(3a+2b)}(1-{k_A}/{k_B}) \hspace{0.2in} e^{i(k_B-k_A)(3a+2b)}(1+{k_A}/{k_B})\\

   \end{array}
   \right ).
\end{eqnarray}
By the above calculation, we can obtain the results of transform
matrixes: \\

(1) For the transform matrix $M_A^1$ of the first period medium
$A$ is independent form. \\

(2) For the transform matrixes $M_A^N$ of the N-th period $(N\geq
2)$, they can be written as
\begin{eqnarray}
M_A^N=\frac{1}{2} \left ( \begin{array}{ll}
  e^{i(k_B-k_A)(N-1)(a+b)}(1+{k_B}/{k_A}) \hspace{0.1in} e^{-i(k_A+k_B)(N-1)(a+b)}(1-{k_B}/{k_A})\\
e^{i(k_A+k_B)(N-1)(a+b)}(1-{k_B}/{k_A}) \hspace{0.1in} e^{i(k_A-k_B)(N-1)(a+b)}(1+{k_B}/{k_A})\\

   \end{array}
   \right ),
 \end{eqnarray}\\

(3) For the transform matrixes $M_B^N$ of the N-th period $(N\geq
1)$, they can be written as
\begin{eqnarray}
M_B^N=\frac{1}{2} \left ( \begin{array}{ll}
  e^{i(k_A-k_B)(N(a+b)-b)}(1+{k_A}/{k_B}) \hspace{0.2in} e^{-i(k_A+k_B)(N(a+b)-b)}(1-{k_A}/{k_B})\\
e^{i(k_A+k_B)(N(a+b)-b)}(1-{k_A}/{k_B}) \hspace{0.2in} e^{i(k_B-k_A)(N(a+b)-b)}(1+{k_A}/{k_B})\\

   \end{array}
   \right ).
\end{eqnarray}
By the quantum transform matrixes, we can give their relations: \\
\begin{figure}[tbp]
\includegraphics[width=16 cm]{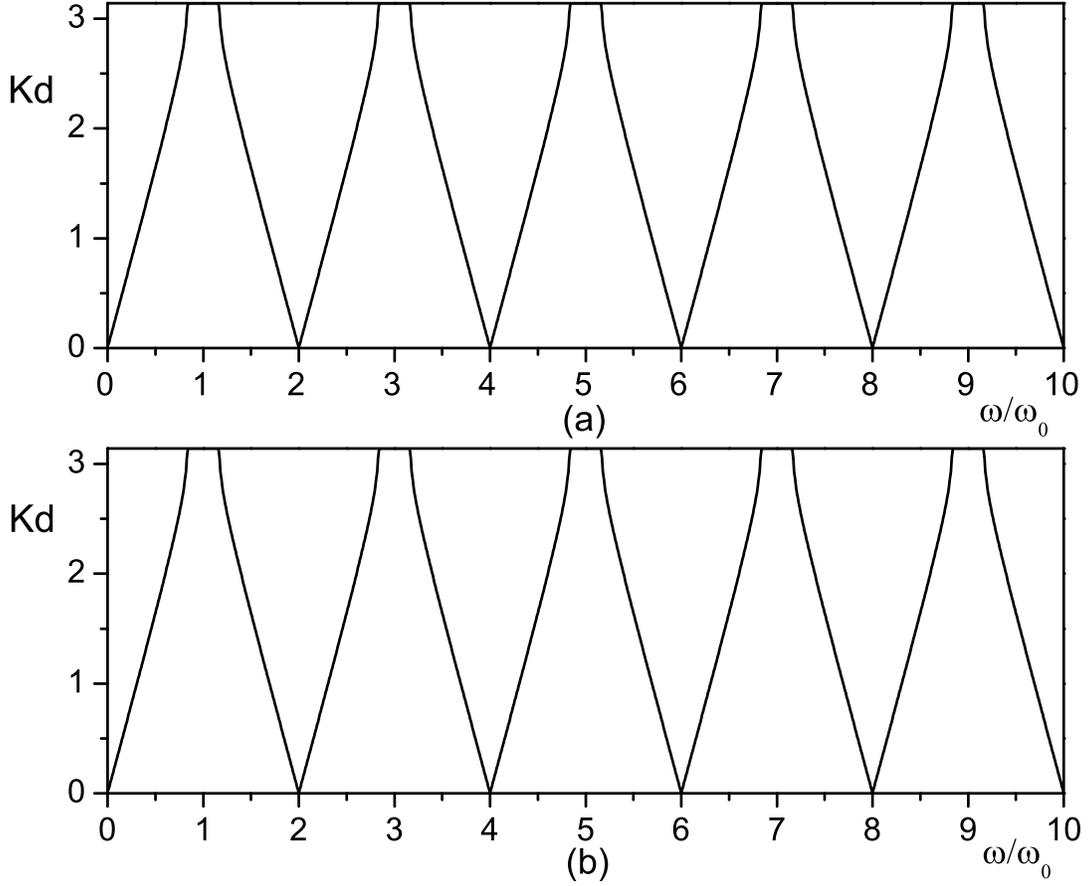}
\caption{comparing quantum dispersion relation (a) with classical
dispersion relation (b)}
\end{figure}
\begin{figure}[tbp]
\includegraphics[width=16 cm]{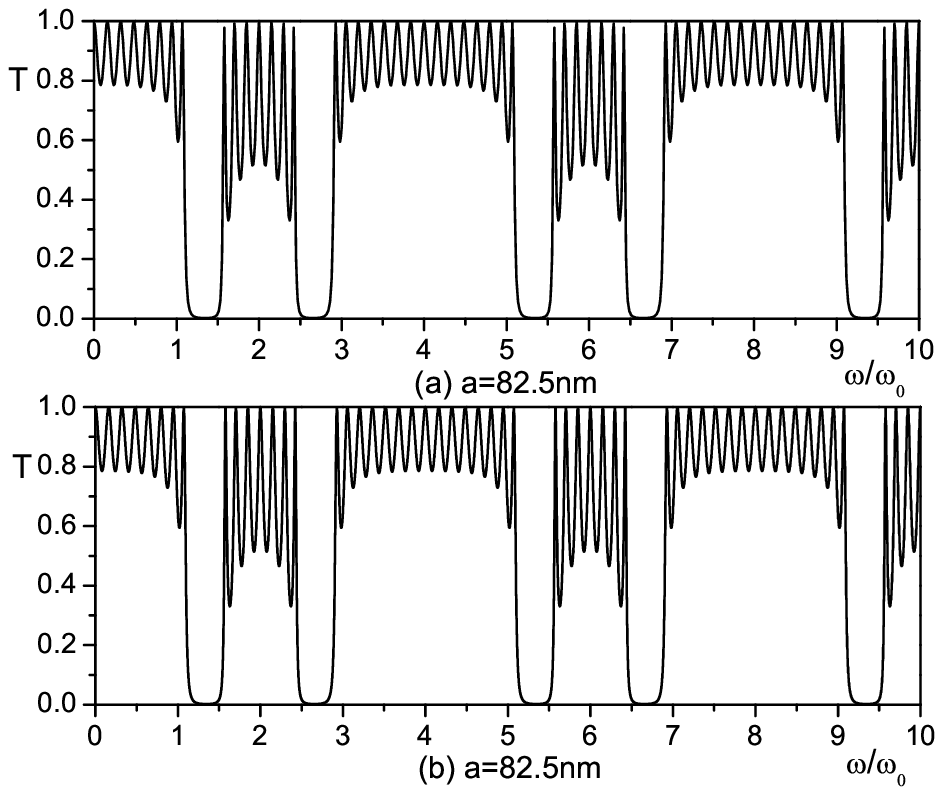}
\caption{comparing quantum transmissivity (a) with classical
 transmissivity (b)}
\end{figure}
\begin{figure}[tbp]
\includegraphics[width=16 cm]{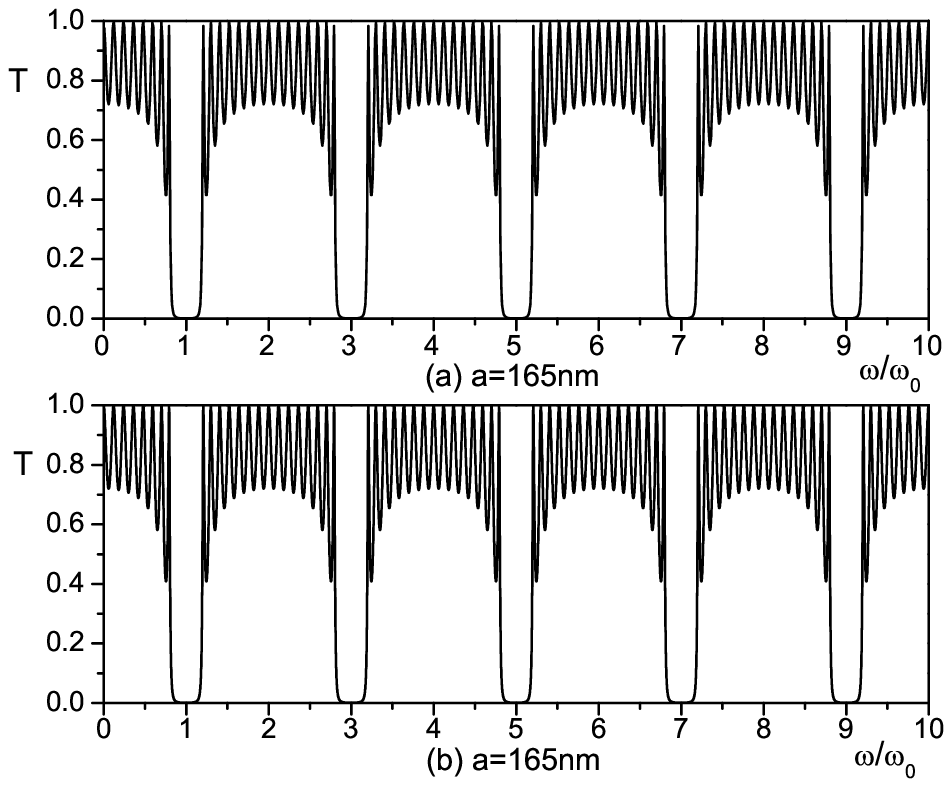}
\caption{comparing quantum transmissivity (a) with classical
 transmissivity (b)}
\end{figure}
\begin{figure}[tbp]
\includegraphics[width=16 cm]{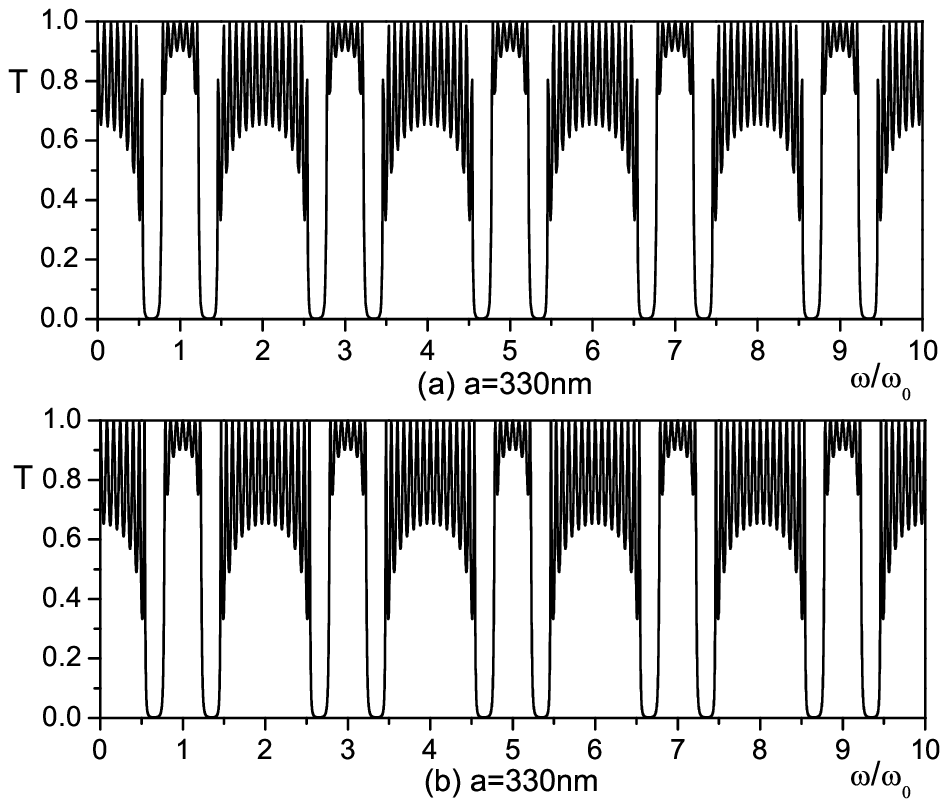}
\caption{comparing quantum transmissivity (a) with classical
 transmissivity (b)}
\end{figure}
\begin{figure}[tbp]
\includegraphics[width=16 cm]{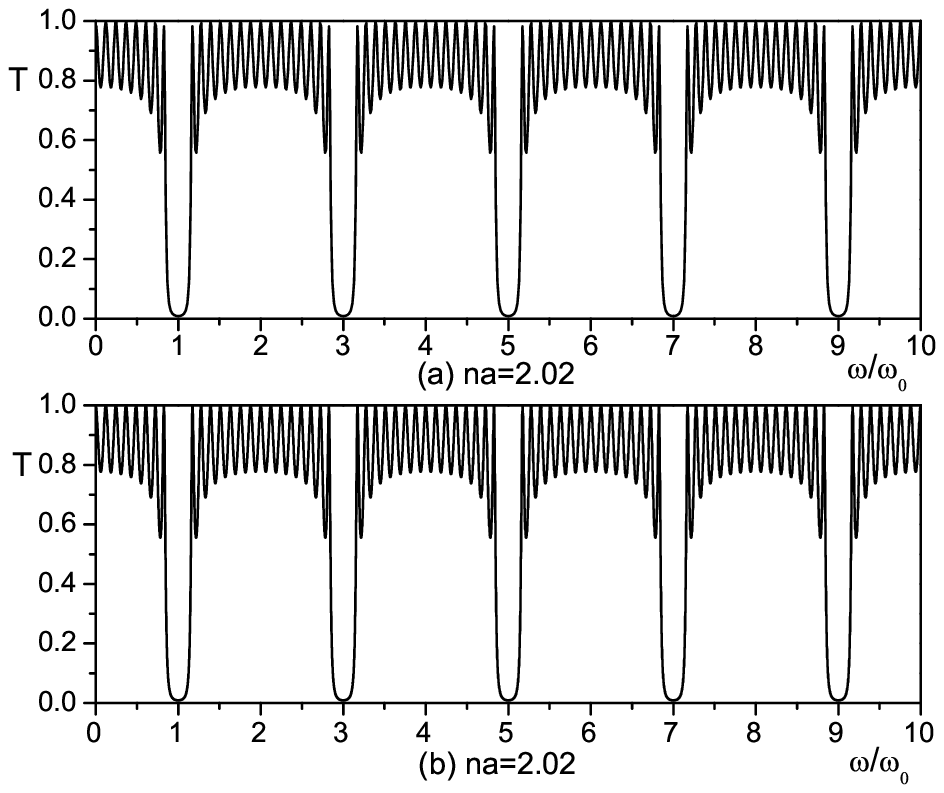}
\caption{comparing quantum transmissivity (a) with classical
 transmissivity (b)}
\end{figure}
\begin{figure}[tbp]
\includegraphics[width=16 cm]{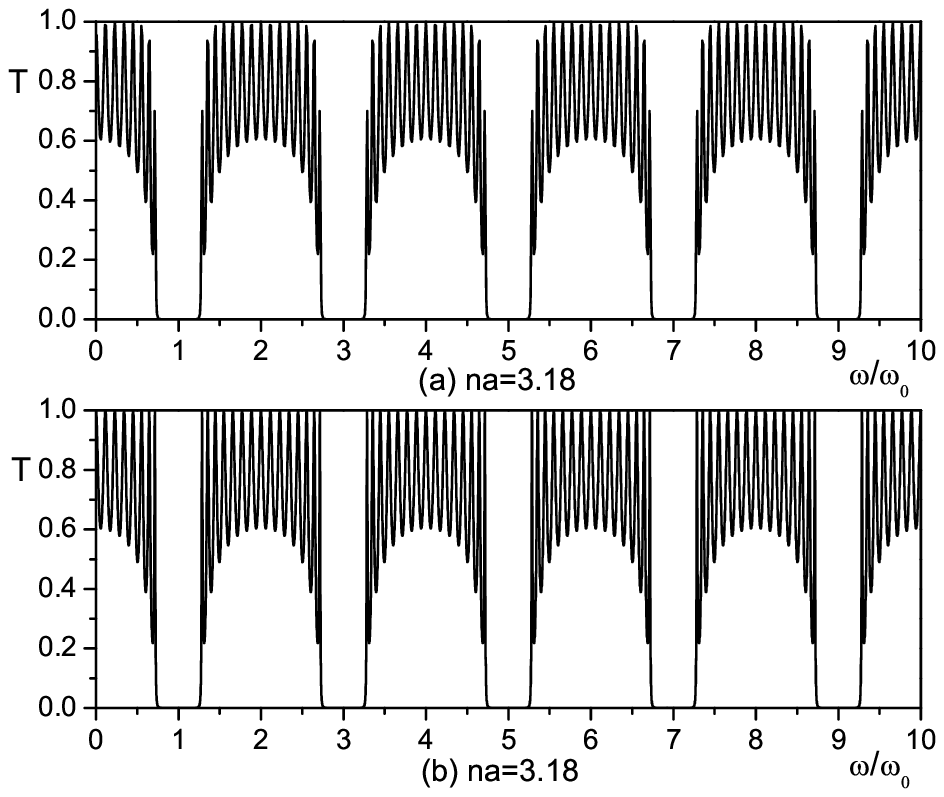}
\caption{comparing quantum transmissivity (a) with classical
 transmissivity (b)}
\end{figure}
(1) The representation of the first period quantum transform
matrixes are
\begin{eqnarray} \left ( \begin{array}{ll}
 A_{k_A}^{1}\\
 A_{-k_A}^{1}\\
   \end{array}
   \right )=M_A^1\left ( \begin{array}{ll}
F\\
F'\\
   \end{array}
   \right ),
\end{eqnarray}
\begin{eqnarray}
\left ( \begin{array}{ll}
B_{k_B}^{1}\\
 B_{-k_B}^{1}\\
   \end{array}
   \right )=M_B^1\left ( \begin{array}{ll}
A_{kA}^1\\
A_{-kA}^1\\
   \end{array}
   \right )=M_B^1M_A^1\left ( \begin{array}{ll}
F\\
F'\\
   \end{array}
   \right )=M^1\left ( \begin{array}{ll}
F\\
F'\\
   \end{array}
   \right ).
\end{eqnarray}\\

(2) The representation of the second period quantum transform
matrixes are
\begin{eqnarray}
\left ( \begin{array}{ll}
A_{k_A}^{2}\\
 A_{-k_A}^{2}\\
   \end{array}
   \right )=M_A^2\left ( \begin{array}{ll}
B_{kB}^1\\
B_{-kB}^1\\
   \end{array}
   \right )=M_A^2M_B^1M_A^1\left ( \begin{array}{ll}
F\\
F'\\
   \end{array}
   \right )=M_A^2M^1\left ( \begin{array}{ll}
F\\
F'\\
   \end{array}
   \right ),
\end{eqnarray}
\begin{eqnarray}
\left ( \begin{array}{ll}
B_{k_B}^{2}\\
 B_{-k_B}^{2}\\
   \end{array}
   \right )=M_B^2\left ( \begin{array}{ll}
A_{kA}^2\\
A_{-kA}^2\\
   \end{array}
   \right )=M_B^2M_A^2M_B^1M_A^1\left ( \begin{array}{ll}
F\\
F'\\
   \end{array}
   \right )=M^2M^1\left ( \begin{array}{ll}
F\\
F'\\
   \end{array}
   \right ).
\end{eqnarray}\\

(3) Similarly, the representation of the N-th period quantum
transform matrixes are
\begin{eqnarray}
\left ( \begin{array}{ll}
A_{k_A}^{N}\\
 A_{-k_A}^{N}\\
   \end{array}
   \right )=M_A^NM_B^{N-1}M_A^{N-1}\cdot\cdot\cdot M_A^2M_B^1M_A^1\left ( \begin{array}{ll}
F\\
F'\\
   \end{array}
   \right )=M_A^NM^{N-1}\cdot\cdot\cdot M^2M^1\left ( \begin{array}{ll}
F\\
F'\\
   \end{array}
   \right ),
\end{eqnarray}
\begin{eqnarray}
\left ( \begin{array}{ll}
B_{k_B}^{N}\\
 B_{-k_B}^{N}\\
   \end{array}
   \right )=M_B^NM_A^NM_B^{N-1}M_A^{N-1}\cdot\cdot\cdot M_A^2M_B^1M_A^1\left ( \begin{array}{ll}
F\\
F'\\
   \end{array}
   \right )=M^NM^{N-1}\cdot\cdot\cdot M^2M^1\left ( \begin{array}{ll}
F\\
F'\\
   \end{array}
   \right )=M\left ( \begin{array}{ll}
F\\
F'\\
   \end{array}
   \right ),
\end{eqnarray}
where
\begin{eqnarray}
M=M^NM^{N-1}\cdot\cdot\cdot M^2M^1=\left ( \begin{array}{ll}
  m_1 \hspace{0.1in} m_2\\
m_3 \hspace{0.1in} m_4\\

   \end{array}
   \right ),
\end{eqnarray}
is the total quantum transform matrix of N period, and
$M^1=M_B^1M_A^1$ is the first period quantum transform matrix,
$M^2=M_B^2M_A^2$ is the second period quantum transform matrix,
and $M^N=M_B^NM_A^N$ is the N-th period
quantum transform matrix.\\
By Eqs. (82) and (83), we can give the wave function of N-th
period in medium $B$, it is
\begin{eqnarray}
{\psi_B^{N}(x)}&=&B_{kB}^{N}e^{ik_{B}x}+B_{-kB}^{N}e^{-ik_{B}x} \nonumber\\
&=&(m_1F+m_2F')e^{ik_{B}x}+(m_3F+m_4F')e^{-ik_{B}x}.
\end{eqnarray}
In FIG. 3, the transmission wave function is
\begin{eqnarray}
\psi_{D}(x)=De^{i K x}.
\end{eqnarray}
At $x=N(a+b)$, by the continuation of wave function and its
derivative, we have
\begin{eqnarray}
(m_1F+m_2F')e^{ik_{B}N(a+b)}+(m_3F+m_4F')e^{-ik_{B}N(a+b)}=De^{iKN(a+b)},
\end{eqnarray}
and
\begin{eqnarray}
\frac{k_B}{K}(m_1F+m_2F')e^{ik_{B}N(a+b)}-\frac{k_B}{K}(m_3F+m_4F')e^{-ik_{B}N(a+b)}=De^{iKN(a+b)},
\end{eqnarray}
we can obtain
\begin{eqnarray}
\frac{F'}{F}=\frac{m_1(K-k_B)e^{ik_{B}N(a+b)}+m_3(K+k_B)e^{-ik_{B}N(a+b)}}{m_2(k_B-K)e^{ik_{B}N(a+b)}-m_4(K+k_B)e^{-ik_{B}N(a+b)}},
\end{eqnarray}
By Eqs. (86)-(88), we have
\begin{eqnarray}
t=\frac{D}{F}=(m_1+m_2\frac{F'}{F})e^{i(k_{B}-K)N(a+b)}+(m_3+m_4\frac{F'}{F})e^{-i(k_{B}+K)N(a+b)},
\end{eqnarray}
and the quantum transmissivity $T$ is
\begin{eqnarray}
T=|t|^2.
\end{eqnarray}

\newpage
\vskip 5pt {\bf 6. Numerical result} \vskip 5pt

In this section, we report our numerical results of quantum
transmissivity and quantum dispersion relation. The main
parameters are: medium $B$ is $MgF_2$, its refractive indexes is
$n_b=1.38$, and its thickness is $b=281nm$. The medium $A$ is
$ZnS$, its refractive indexes is $n_a=2.35$, and its thickness is
$a=165nm$. The central frequency is $\omega_0=271THz$, and the
periodicity $N=8$. In numerical calculation, we compare quantum
dispersion relation and quantum transmissivity with classical
dispersion relation and transmissivity. With Eq. (45), we can
investigate the quantum dispersion relation, and compare it with
classical dispersion relation, which are shown in FIG. 4. The FIG.
4 (a) and (b) are quantum dispersion relation and classical
dispersion relation, respectively. We can find the dispersion
relation of classical and quantum are identical. With Eqs. (89)
and (90), we can calculate the quantum transmissivity, and compare
it with classical transmissivity. Firstly, we study the effect of
thickness $a$ on the quantum and classical transmissivity, which
are shown in FIGs. 5, 6 and 7 according to thickness $a$ are
$82.5nm$, $165nm$ and $330nm$, respectively. We can find when the
thickness $a$ increase the band gaps width decrease and the number
of band gaps increase for the quantum and classical
transmissivity, and also find the quantum and classical
transmissivity are identical. Then, we study the effect of
refractive indexes $n_a$ on the quantum and classical
transmissivity, which are shown in FIGs. 8 and 9 according to
refractive indexes $n_a$ are $2.02$ and $3.18$, respectively. We
can find when the refractive indexes $n_a$ increase the band gaps
width increase and the number of band gaps invariant for the
quantum and classical transmissivity, and also find the quantum
and classical transmissivity are identical.

\newpage
 \vskip 5pt
{\bf 5. Conclusion} \vskip 5pt

In summary, we have firstly presented a new quantum theory to
study one-dimensional photonic crystals. We give quantum
dispersion relation and quantum transmissivity, and compare them
with the classical dispersion relation and classical
transmissivity. By the calculation, we find the classical and
quantum dispersion relation and transmissivity are identical. The
new approach we can be studied two-dimensional and
three-dimensional photonic crystals.
 
\end{document}